\documentclass[letterpaper, 10 pt, conference]{ieeeconf}

\IEEEoverridecommandlockouts                              

\usepackage{graphics} 
\usepackage{epsfig} 
\usepackage{mathptmx} 
\usepackage{times} 
\usepackage{cite}
\usepackage{textcomp}
\usepackage{xcolor}
\usepackage{soul}
\usepackage{amsmath}
\usepackage{amssymb}
\usepackage{amsfonts}
\usepackage{graphicx}
\usepackage{algorithm}
\usepackage{algcompatible}
\usepackage{graphicx}
\usepackage{subfig}

\usepackage[top=0.75in,bottom=0.8in,right=0.75in,left=0.75in]{geometry}

\def\BibTeX{{\rm B\kern-.05em{\sc i\kern-.025em b}\kern-.08em
    T\kern-.1667em\lower.7ex\hbox{E}\kern-.125emX}}

\newcommand{\norm}[2]{\left \lVert #1 \right \rVert_{#2}}
\algnewcommand\algorithmicreturn{\textbf{return}}
\algnewcommand\RETURN{\State \algorithmicreturn}%
\algrenewcommand\algorithmicrequire{\textbf{Input:}}
\algrenewcommand\algorithmicensure{\textbf{Output:}}
\newtheorem{example}{Example}

\title{\LARGE \bf
Collision-Free Multi-Agent Coverage Control for Non-Cooperating Swarms: Preliminary Results
}

\author{Karolina Schmidt and Luis Rodrigues*
\thanks{*Karolina Schmidt and Luis Rodrigues are with the Department of Electrical and Computer Engineering, Concordia University,
Montreal, QC H3G 2W1, Canada}%
}

\begin{document}

\maketitle
\thispagestyle{empty}
\pagestyle{empty}

\begin{abstract}

The main contribution of this paper is a methodology for multiple non-cooperating swarms of unmanned aerial vehicles to independently cover a common area. In contrast to previous research on coverage control involving more than one swarm, this paper does not assume cooperation between distinct groups but considers them as entirely independent units following their own objectives. Using Voronoi tesselation, collision-free motion of agents within the same swarm has been proved before. However, as is shown in Example 1 of this paper, in the case of multiple swarms with inter-swarm but without intra-swarm collaboration, these guarantees do not hold. We address this issue by proposing an algorithm to achieve maximum coverage with multiple swarms while avoiding collisions between agents. Thus, the Optimal Reciprocal Collision Avoidance method used for safe navigation in multi-agent scenarios is adapted to suit the needs of Voronoi-based coverage control with more than one swarm. The functionality of the proposed technique is validated through Monte Carlo simulations.

\end{abstract}

\section{Introduction} \label{introduction}

Recent events of wildfire in Canada as well as other countries worldwide have risen concerns all around the globe. Fires do not only pose hazards to populations but also to firefighting teams. A strong need for reliable means of assistance arises to ensure security for communities as well as individuals while monitoring affected areas. In that sense, the use of Unmanned Aerial Vehicles (UAV) has gained importance in recent years for search and rescue missions as well as fire fighting and monitoring. In this context, it is common for UAVs to work in swarms aiming to collaboratively pursue a mutual goal. A prevalent example for such scenarios is the provision of maximum coverage over an area of interest with a group of multiple agents. Furthermore, applications demonstrating similar needs not only comprise planetary exploration with multiple swarms of vehicles competing to investigate the same area, but also drones belonging to different companies providing mobile coverage in remote areas. The problem of coverage control incorporates the planning of motion for each vehicle of the multi-agent system to conjointly achieve a common objective. 

Coverage control has been a widely studied subject during the past two decades. Using Lloyd's iterative algorithm \cite{LloydsAlgorithm}, Cortés et al. \cite{CortesMartinez1,CortesMartinez2} introduced the use of centroidal Voronoi partitions and proximity graphs for coverage control. Gradient descent algorithms to coordinate a group of agents are presented and guarantees for the convergence of the algorithms are studied. Applications of centroidal Voronoi tesselations are discussed in \cite{DuEtAl1999}. The coverage problem is formulated as an optimal control problem considering energy efficiency and conservation in \cite{MoarrefRodrigues2014}. Reference \cite{NguyenRodriguesManiuOlaru2016} addresses the same problem for discrete-time systems. Analytic expressions for the rate of change of the mass and the location of the center of mass of a Voronoi cell are presented in \cite{DipernaRodrigues2017}. A common feature in the majority of published research in coverage control is the consideration of one single swarm of agents working as a team towards a common goal. However, multiple applications require several groups of UAVs to independently provide maximum coverage over the same area for various purposes, e.g. search and rescue and fire monitoring. Although in \cite{HaghighiCheah2012} more than one swarm of vehicles is incorporated into a multi-agent scenario, the objective is multi-agent formation rather than coverage control. Two years later, Sharifi investigated coverage control using a group of unmanned ground vehicles along with a swarm of UAVs in \cite{SharifiPhDThesis2014}. Yet, the two swarms collaborate with each other to cover the given area as a unit. In \cite{AtınçEtAl2020}, the authors examined the use of more than one swarm of vehicles to cover larger areas in shorter time adopting a leader-follower approach. Similarly to reference \cite{SharifiPhDThesis2014}, different swarms jointly work towards a common goal. The division of an overall unit of robots working towards one common goal into sub-groups of similar size to increase regional coverage speed and reduce the moving distances of the agents is proposed in \cite{WangEtAl2021}.

As a first approach to ensure collision-free motion for agents of all swarms, the Velocity Obstacles (VO) method and its extensions, with a primary focus on Optimal Reciprocal Collision Avoidance (ORCA), are adapted in this work. A detailed description of the ORCA method is given in section \ref{orca}. The VO technique was first formalized in 1998 by Fiorini and Shiller \cite{VelocityObstacles1998} who introduced the selection of maneuvers to avoid static as well as moving obstacles based on operations within the velocity space. Assuming obstacles with constant velocities and directions of motion, the agents select their velocity outside the Velocity Obstacle ($\mathcal{VO}$), which is defined as the subset of the velocity space that results in collision at a future time. Reference \cite{VelocityObstaclesReview2024} provides an overview of the VO method and its extensions. 

An extension of VO for multi-agent navigation considering reactive behavior of other agents under the assumption of similar avoidance strategies was proposed by Van Den Berg et al. in \cite{ReciprocalVelocityObstacles2008}. Further research on VO for multi-agent navigation includes the work of Guy et al. \cite{ClearPathFVO2009} and Van Den Berg et al. \cite{ORCA2011}. The former introduced truncated collision cones and formulated an optimization problem to extend the VO approach and ensure collision-free multi-agent motion. Furthermore, they include considerations for computational efficiency through data and thread-level parallelism. The authors of \cite{ORCA2011} adopted the idea of VO in the form of truncated cones for collision avoidance in multi-agent systems with independently operating robots. Collision-free motion is achieved by deriving half-spaces of permitted velocities so that a low-dimensional linear program can be solved to guarantee local collision avoidance.

The main contribution of this paper is a collision-free algorithm for multiple non-cooperating swarms of UAVs to independently cover a common area. To the best of the author's knowledge, it is the first investigation presenting coverage control of several non-collaborating multi-agent swarms operating within the same space. In contrast to previous work, distinct swarms act as entirely independent units without explicitly exchanging information. Besides the possibility of agents encountering each other while moving towards their desired positions, there exist cases where distinct swarms of vehicles aim to converge to the same configurations. Such a scenario is presented in example 1. Therefore, to avoid collisions between agents the ORCA method is adapted to suit Voronoi-based coverage control. Monte Carlo simulations validate the proposed algorithm.

The structure of this paper is as follows. Section \ref{preliminaries} introduces preliminary notions and definitions on Voronoi tesselation (section \ref{voronoi}), coverage control (section \ref{coverage control}), a motivating example (section \ref{example1}), and collision-free navigation using ORCA (section \ref{orca}). The proposed extension of coverage control to independent swarms of agents safely operating within a common area is introduced in section \ref{proposed method}. Section \ref{simulation} presents Monte Carlo simulations for a specific example. Conclusions follow in section \ref{conclusions}.

\section{Preliminaries} \label{preliminaries}
This section summarizes underlying notions as well as definitions introduced in the related literature and presents a motivating example. After a brief description of Voronoi tesselation in subsection \ref{voronoi}, its incorporation into coverage control is described in subsection \ref{coverage control} following \cite{CortesMartinez1,CortesMartinez2,MoarrefRodrigues2014}. An example illustrating the particular importance of collision avoidance in coverage control is given in subsection \ref{example1}. Subsection \ref{orca} presents an introduction to the method of ORCA as introduced in \cite{ORCA2011}.

\subsection{Voronoi Tesselation} \label{voronoi}

\begin{figure}[t]
    \centering
    \includegraphics[scale=0.5]{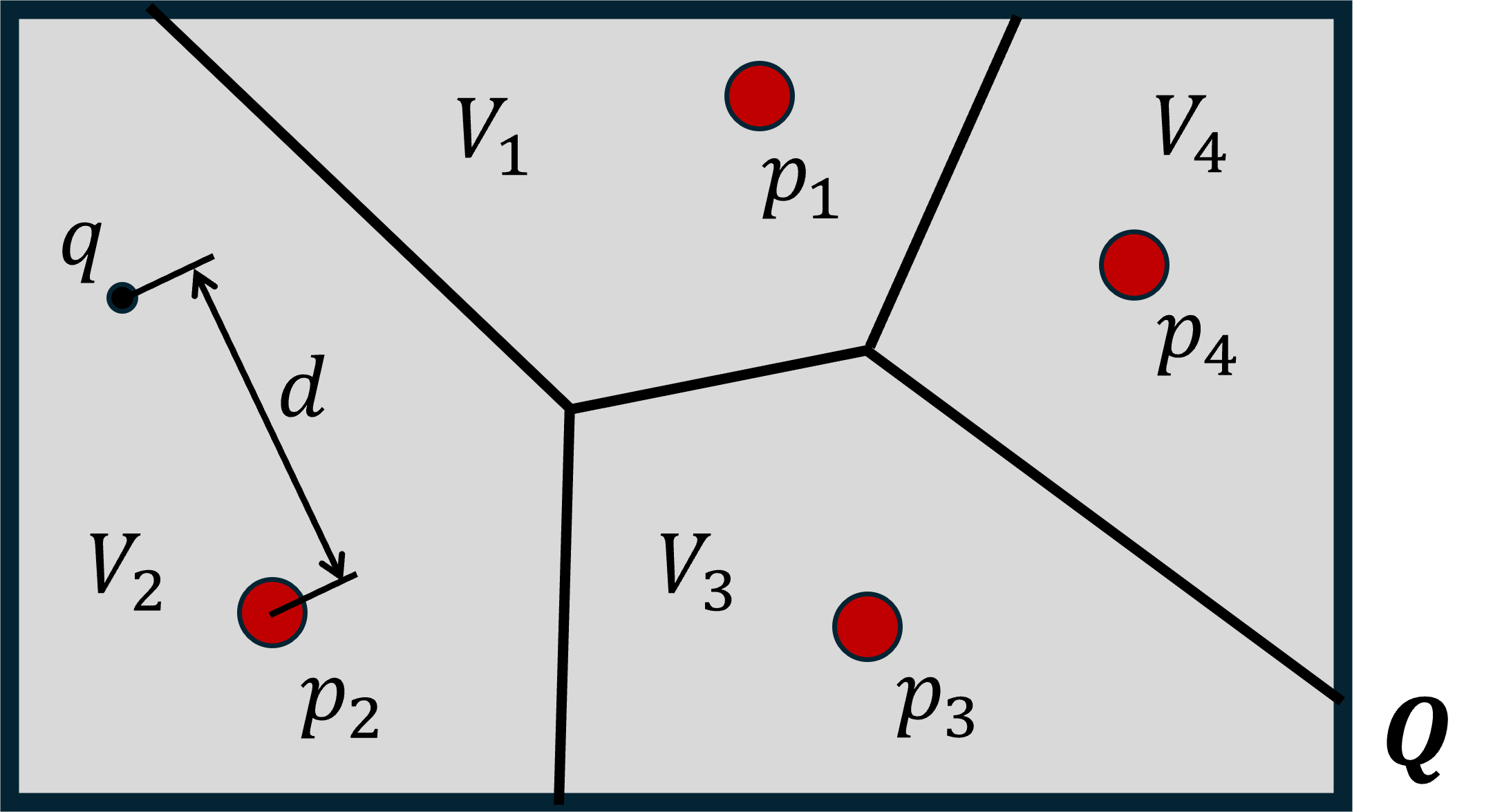}
    \caption{Voronoi tesselation of an area $Q$ containing four generators}
    \label{fig:voronoidiagram}
\end{figure}

Consider $n$ distinct points referred to as generators located within a convex area $Q$. Voronoi tesselation is obtained by partitioning $Q$ into subsets, each taking the shape of a convex polytope, with the Lebesgue measure of overlaps between subsets being zero. For example, the boundary between Voronoi cells in 2D corresponding to neighboring generators is therefore defined by a straight line with each point on the line being equidistant from the two locations of the respective generators. Each generator $p_i, i \in \{ 1,...,n \}$ is associated with one subset called Voronoi cell $V_i(P), i \in \{ 1,...,n \}$, where $P=\{ p_1,...,p_n \} \subset Q$. The division into these Voronoi cells is realized so that the distance from any point $q \in Q$ to the generator of its Voronoi cell is less than or equal to the distance to any other generator $p_j$: $V_i(P)= \{ q \in Q \, | \norm{q-p_i}{2} \leq \norm{q-p_j}{2}\forall p_j \in P \}$. A Voronoi tesselation with the generators being located at the center of mass of their Voronoi cells is called a centroidal Voronoi configuration. Figure \ref{fig:voronoidiagram} shows an example of a Voronoi diagram for an area containing four generators.

\subsection{Multi-Agent Coverage Control using Voronoi Tesselation} \label{coverage control}

One field of application of Voronoi tesselation is multi-agent coverage control. Multi-agent coverage control deals with the problem of deploying a number $n$ of collaborating agents whose positions are denoted by $x_i, i \in \{ 1,...,n \}$ so that maximum coverage of a given area $Q$ is achieved. Voronoi tesselation is obtained with each agent $i$ acting as a generator for their respective Voronoi cell $V_i$. Coverage at a point $q \in Q$ is then inversely proportional to the square of the distance from $q$ to the agent located closest to $q$, here denoted by $i(q): Q \rightarrow \{ 1,...,n \}$. A performance function is defined as $f(x_{i(q)},q)= ||x_{i(q)}-q||^2=d^2$, where $d$ is the distance between $q$ and the corresponding agent $i(q)$. Low values of $f(x_{i(q)},q)$ indicate a high level of coverage at $q$. Furthermore, a density function $\phi (q)$ contains the priority of coverage at any point $q$ within the area $Q$. The candidate Lyapunov function to be minimized in order to maximize overall coverage over the area of interest is therefore \cite{MoarrefRodrigues2014}

\begin{figure*} [ht]
    \centering
    \subfloat[]{\includegraphics[width=0.33\textwidth]{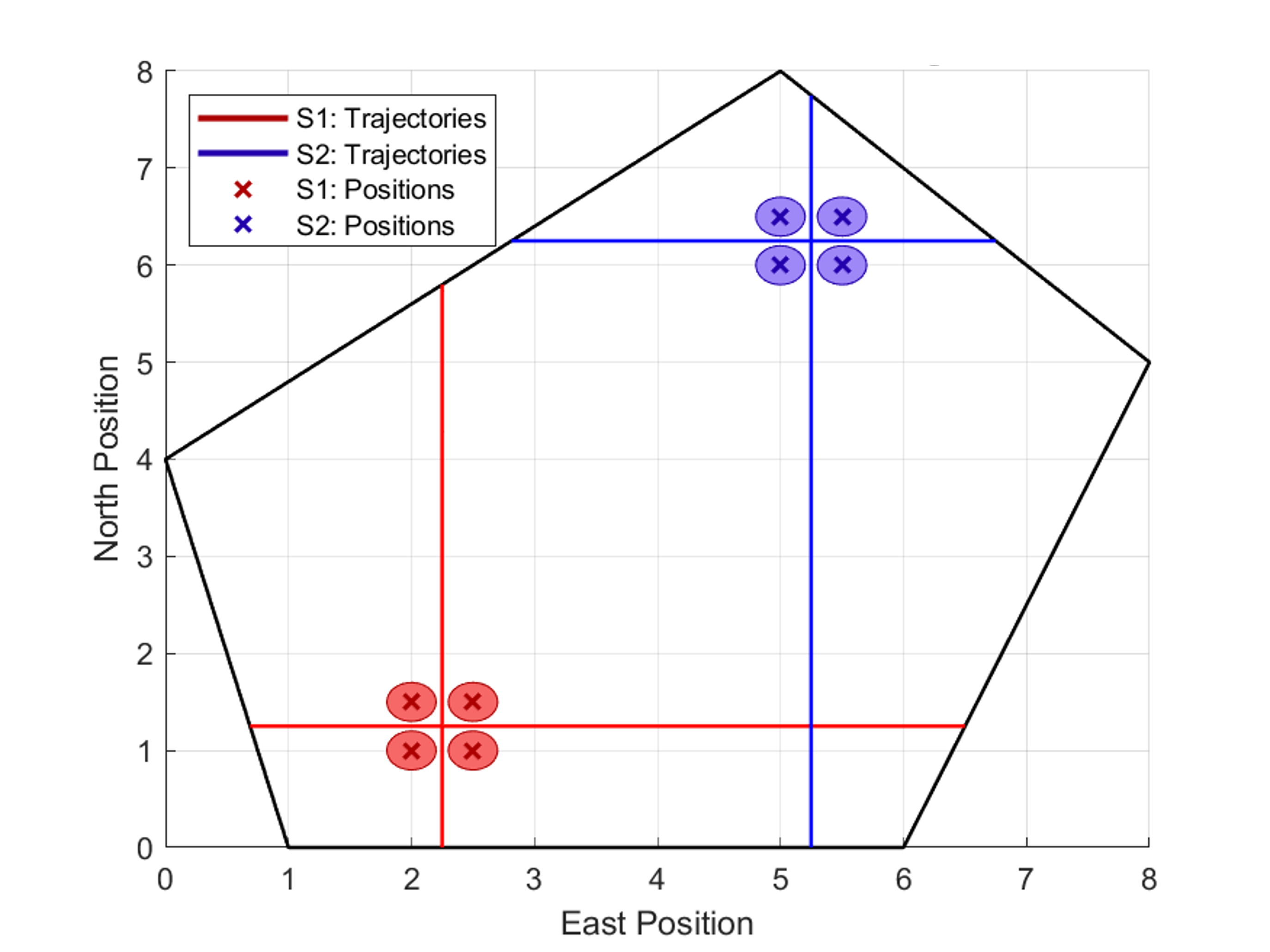}} 
    \subfloat[]{\includegraphics[width=0.33\textwidth]{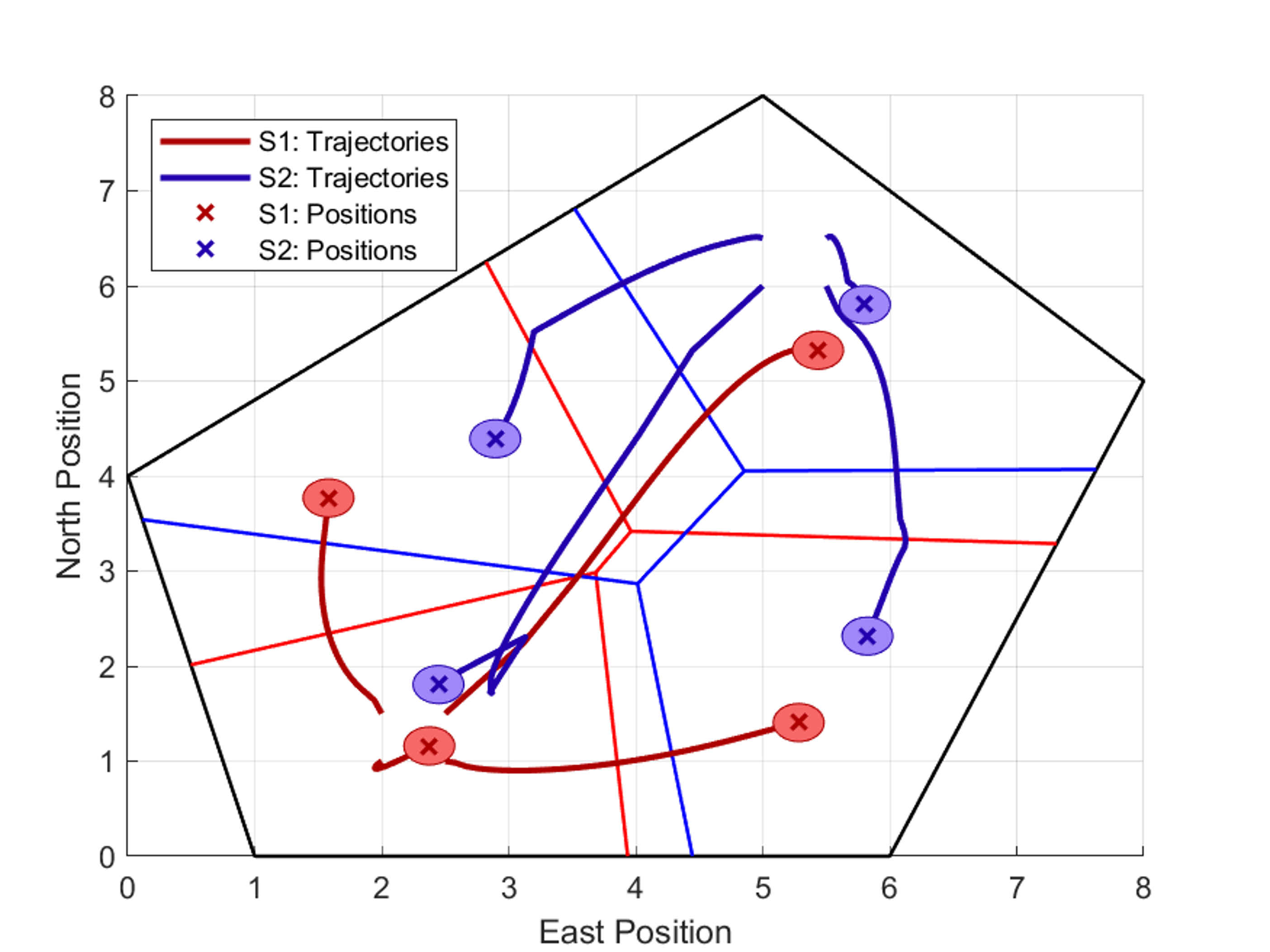}}
    \subfloat[]{\includegraphics[width=0.33\textwidth]{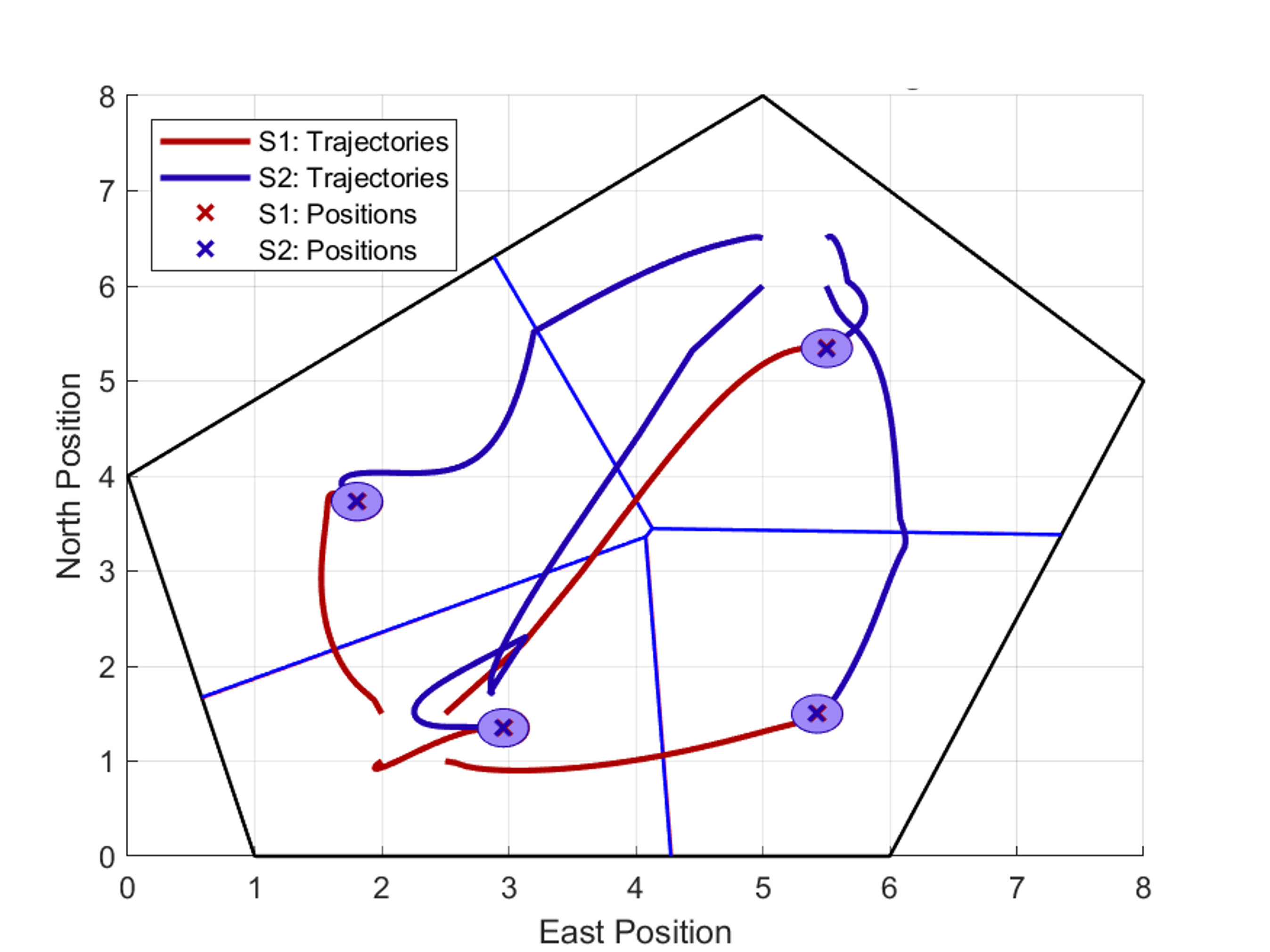}}
    
    \caption{Simulation of coverage control with two non-cooperating swarms without collision avoidance showing the trajectories and final positions of the agents of swarm 1 (S1) and swarm 2 (S2) (a) at the first iteration (b) after 50 iterations (c) after convergence to their final positions. Notice that the red agents are hidden behind the blue ones.}
    \label{fig:motivatingexample}
\end{figure*}

\begin{equation} \label{functiontominimize}
     V(x)= \int_{Q} f(x_{i(q)},q) \phi (q) dq
\end{equation}
and indicates the poorness of coverage over $Q$, with $x=[x_1^T,...,x_n^T]^T$ containing the positions of the agents. By rewriting (\ref{functiontominimize}) considering all Voronoi cells $V_i$ individually instead of $Q$ as a whole and summing up the results, one obtains
 \begin{equation}
     V(x) = \sum_{i=1}^{n} \int_{V_i} f(x_{i(q)},q) \phi (q) dq.
 \end{equation}
Replacing the performance function yields

 \begin{equation}
     V(x) = \sum_{i=1}^{n} \int_{V_i} ||x_{i(q)}-q||^2 \phi (q) dq.
 \end{equation}
The necessary condition for a local minimum of $V(x)$ is then obtained by differentiating with respect to agent $j$ and can be written as

 \begin{equation} \label{necessarycondition}
     \begin{split}
         \frac{\partial V}{\partial x_j} & = 2 \int_{V_j} (x_j-q)^T \phi (q)dq \\
         & = 2 \left( \int_{V_j} \phi (q) dq \right) \left( x_j- \frac{\int_{V_j}q \phi (q)dq}{\int_{V_j} \phi (q)dq} \right)^T \\
         & = 0.
     \end{split}
 \end{equation}
Note that equation (\ref{necessarycondition}) contains a term for the mass $M_{V_j}$ of Voronoi cell $V_j$:

\begin{equation} \label{mass}
    M_{V_j} = \int_{V_j} \phi (q) dq,
\end{equation}
as well as a term for its center of mass $CM_{V_j}$:

\begin{equation} \label{centerofmass}
    CM_{V_j} = \frac{\int_{V_j}q \phi (q)dq}{\int_{V_j} \phi (q)dq} = \frac{\int_{V_j}q \phi (q)dq}{M_{V_j}}.
\end{equation}
Equation (\ref{necessarycondition}) can therefore be rewritten as

\begin{equation} \label{necessarycondition_short}
    \frac{\partial V}{\partial x_j} = 2M_{V_j} (x_j-CM_{V_j})^T = 0.
\end{equation}
Clearly, this condition is true when $x_j=CM_{V_j}$ which indicates that agent $j$ is located at the center of mass of its Voronoi cell $V_j$. With the mass always being positive, the sufficient condition for a local minimum of $V$ is satisfied,

\begin{equation} \label{sufficientcondition}
    \frac{\partial ^2 V}{\partial x_j^2} = 2M_{V_j} > 0.
\end{equation}

To move each agent towards the center of mass of its Voronoi cell, a commonly used strategy is Lloyd's algorithm \cite{LloydsAlgorithm}. Iteratively, agents compute the center of mass of their Voronoi cells and select velocity vectors pointing in the respective direction. A first order dynamic system is thus

\begin{equation} \label{dynamicsystem}
    \begin{split}
        & \dot{x_j} = u_j, \\
        & u_j = c_j(CM_{V_j}-x_j),
    \end{split}
\end{equation}
where $u_j$ is the control input and $c_j > 0$. Note that agents slow down as they get closer to their goal position.

\subsection{Motivating Example} \label{example1}

Previous work \cite{PiersonEtAl2017} proved that for a single swarm of multiple agents following the Voronoi partitioning approach with the agents represented as points moving towards the center of mass of their respective Voronoi cells to cover a convex area, no collisions between agents occur. This is due to the fact that for a convex area all Voronoi cells take a convex polygonal shape which results in the center of mass of a cell always being within its boundaries. Thus, an agent directly applying its desired control input to move towards the center of mass of its Voronoi cell always stays within the respective boundaries and does not enter the cells corresponding to other agents. 

When multiple non-cooperating swarms of agents operate within a common area to provide coverage, these guarantees for collision avoidance do not hold. Since Voronoi cells of agents belonging to different swarms overlap, collisions with vehicles from other swarms can occur, even if the agents do not leave the boundaries of their cells. A risk of collisions is therefore present during the entire time of travel. Furthermore, with the approach introduced in subsections \ref{voronoi} and \ref{coverage control} it is possible that vehicles belonging to different swarms aim to take identical final locations. One scenario where this is the case is shown in the following example.

\begin{example} \label{ex:example1}
A Matlab simulation demonstrating an example of the coverage task executed by distinct swarms with different initial positions within a common area is performed. The density distribution is uniform and no collision avoidance is incorporated. The simulation shows a case where agents find themselves at identical final locations. Two swarms each containing a number of four agents are initially placed at the following positions:
 \begin{itemize}
     \item swarm 1 (S1): $[(2,1),(2,1.5),(2.5,1),(2.5,1.5)]$,
     \item swarm 2 (S2): $[(5,6),(5,6.5),(5.5,6),(5.5,6.5)]$.
 \end{itemize}
 Both groups independently aim to cover an area defined by a convex polygon with vertices at locations $[(1,0),(6,0),(8,5),(5,8),(0,4)]$. All agents have radii of $0.2$ units. 

Figure \ref{fig:motivatingexample} shows the initial positions of the agents (figure \ref{fig:motivatingexample}a), as well as their positions and traveled trajectories after 50 iterations (figure \ref{fig:motivatingexample}b), and their final positions and overall trajectories after convergence (figure \ref{fig:motivatingexample}c). The first swarm is represented in red while the second swarm is displayed in blue. Notice that in figure \ref{fig:motivatingexample}c the positions of agents from the two swarms coincide, and therefore the red agents are hidden behind the blue ones. The convergence of both swarms to the same local optimum is clearly visible. This highlights the need for reliable collision avoidance.

\end{example}

\subsection{Optimal Reciprocal Collision Avoidance (ORCA)} \label{orca}

Based on the original VO method \cite{VelocityObstacles1998}, ORCA \cite{ORCA2011} is a technique used to modify a desired control input velocity so that no collisions with static and moving obstacles occur within a given time horizon $\tau$. Assuming known positions $x_A$ and $x_B$ as well as known circular or convex polygonal shapes of both a robot $A$ and an obstacle $B$, $A$ computes a truncated cone $\mathcal{VO}_{AB}^{\tau}$ in the velocity space containing all relative velocities of $A$ with respect to $B$ that result in collision within time horizon $\tau$. Given velocities $v_A$ and $v_B$, choosing a new velocity $v_A^{new}$ outside $\mathcal{VO}_{AB}^{\tau}$, leads to $A$ avoiding collisions with $B$ until time $\tau$. Designed for multi-agent scenarios, ORCA assumes shared responsibility between two agents to bypass each other. Therefore, both agents compute a collision cone with respect to one another and adapt their velocities to navigate safely.

\begin{figure}[t]
    \centering
    \includegraphics[scale=0.45]{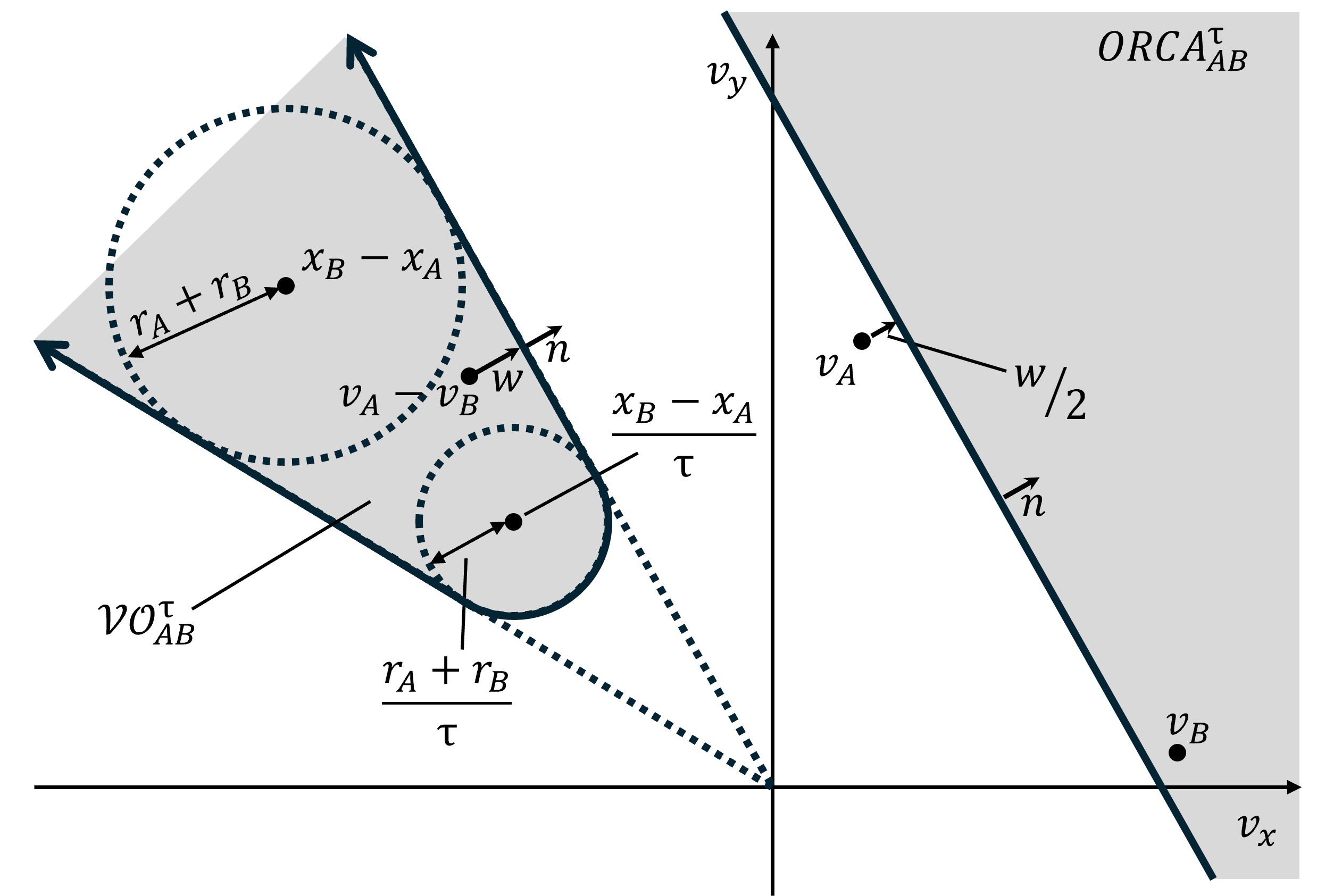}
    \caption{Geometrical interpretation of $\mathcal{VO}_{AB}^{\tau}$ and half-plane $ORCA_{AB}^{\tau}$}
    \label{fig:orca}
\end{figure}

The computation of $\mathcal{VO}_{AB}^{\tau}$ proceeds as follows. Assuming two agents $A$ and $B$ with circular shapes and denoting their radii as $r_A$ and $r_B$, respectively, the $\mathcal{VO}$ for $A$ induced by $B$ for the time window $\tau$ is \cite{ORCA2011}

\begin{equation} \label{velocityobstacle}
    \mathcal{VO}_{AB}^{\tau} = \{ v \, | \, \exists t \in [0, \tau ]::tv \in D(x_B-x_A,r_A+r_B)\}.
\end{equation}
A geometric interpretation of $\mathcal{VO}_{AB}^{\tau}$ is depicted in figure \ref{fig:orca}. Note that $\mathcal{VO}_{BA}^{\tau}$ is the respective collision cone for agent $B$ induced by $A$ which is symmetric to $\mathcal{VO}_{AB}^{\tau}$ at the origin. In expression (\ref{velocityobstacle}), $D$ is an open disc of radius $r=r_A+r_B$ centered at $x=x_B-x_A$, thus

\begin{equation} \label{disc}
    D(x,r)= \{ q \, | \, ||q-x|| < r \}.
\end{equation}

If the relative velocity of $A$ with respect to $B$ falls within $\mathcal{VO}_{AB}^{\tau}$ a collision will occur before time $\tau$ in case both $A$ and $B$ maintain their velocities $v_A$ and $v_B$. To avoid that collision, agent $A$ computes the vector $w$ from $v_{AB}=v_A-v_B$ to the closest point on the boundary $\partial \mathcal{VO}_{AB}^{\tau}$ of Velocity Obstacle $\mathcal{VO}_{AB}^{\tau}$ as \cite{ORCA2011}

\begin{equation} \label{vectortoclosestpointonboundary}
    w = ( \underset{v \in \partial \mathcal{VO}_{AB}^{\tau}}{\text{argmin}} ||v-v_{AB}||) - v_{AB}.
\end{equation}

Computing the outward normal vector $n$ on $\partial \mathcal{VO}_{AB}^{\tau}$ at $v_{AB}+w$, a half-plane of permitted velocities for agent $A$ pointing in direction $n$ can be inferred. Starting at point $v_A+ \frac{1}{2} w$, it is defined as \cite{ORCA2011}

\begin{equation} \label{orca_halfplane}
    ORCA_{AB}^{\tau} = \{ v \, | \, (v-(v_A+ \frac{1}{2} w)) \cdot n \geq 0 \}
\end{equation}
and depicted in figure \ref{fig:orca}. Note that this is only true under the assumption of shared responsibility of the agents to avoid each other. Agent $B$ computes its half-plane containing the respective permitted velocities in a similar manner. To avoid collisions with a number $m$ of agents, agent $A$ computes one half-plane for each of the other agents $B_1,...,B_m$ independently. Moreover, a disc $D(0,v_A^{max})$ restricts the magnitude of the new velocity vector $v_{new}$ to the maximum speed of agent $A$ denoted by $v_A^{max}$. The resulting set of permitted velocities $ORCA_{A}^{\tau}$ for agent $A$ is then the intersection of all half-planes and $D(0,v_A^{max})$:

\begin{equation} \label{intersectionpermittedvelocities}
    ORCA_{A}^{\tau} = D(0,v_A^{max}) \cap \bigcap_{i=1}^{m} ORCA_{AB_i}^{\tau}.
\end{equation}

Finally, the desired velocity vector $v_A^{pref}$ of agent $A$ is corrected to its new velocity: 

\begin{equation} \label{newvelocity}
    v_A^{new} = \underset{v \in ORCA_{A}^{\tau}}{\text{argmin}} \norm{v-v_A^{pref}}{}.
\end{equation}

\section{Proposed Algorithm} \label{proposed method}

In this section, we propose an algorithm (algorithm \ref{algorithm:v_new}) to incorporate collision avoidance into Voronoi-based coverage control for the case where multiple non-cooperating swarms independently aim to cover a shared two-dimensional space. The following assumptions are made:

\begin{enumerate}
    \item all swarms aim to fulfill the coverage task within an area of uniform density distribution,
    \item all swarms follow the same approach to cover the area and avoid other vehicles,
    \item group affiliation as well as circular shape is known,
    \item knowledge regarding positions of other agents can be obtained through communication between vehicles within the same team along with sensors to determine the locations of agents from other swarms.
\end{enumerate}

Note that although this work assumes circular shape of the agents, other shapes can be incorporated.

Algorithm \ref{algorithm:v_new} adapts ORCA and incorporates it into Voronoi-based coverage control to avoid collisions between vehicles belonging to $N$ different swarms. For agents with first-order dynamics frequently considered in coverage control and introduced in section \ref{coverage control}, the control input is a velocity vector that an agent $i$ aims to adopt: $v_{i}^{pref}=u_i$. With ORCA operating in the velocity space, the algorithm uses this method to modify the desired control inputs of the agents and obtain a new collision-free velocity $v_i^{new}$.

The algorithm is executed by each agent individually at each iteration. The notation of $x_{i,j}$ and $v_{i,j}$ indicates the position and velocity of agent $i$ belonging to swarm $j$. $X$ stands for a set containing the positions of all agents, whereas $X_j \subset X$ includes only those belonging to swarm $j$. After updating the positions of all agents as well as its own for the current iteration (lines 1,2 in algorithm \ref{algorithm:v_new}), the agent computes the Voronoi tesselation generated by the members of its swarm and calculates the location of the center of mass of its Voronoi cell (lines 3,4). Subsequently (line 5), the computation of its desired control input velocity follows, initially without accounting for collisions with other agents following equation (\ref{preferredinputvelocity}), 

\begin{equation} \label{preferredinputvelocity}
    v_i^{pref} = u_i = c_i(CM_{V_i}-x_i).
\end{equation}

Corrections to the preferred velocity to avoid collisions with agents from the same as well as other swarms are made using ORCA. For each swarm, a Voronoi tesselation over the area of interest is computed along with the centers of mass and estimates of the desired velocities of the respective agents making use of assumption 2 (lines 8-10).

The truncated $\mathcal{VO}$-cones with respect to the other agents are then computed and the corresponding half-planes are inferred to obtain the set of permitted velocities (lines 11-16). Finally, the agent chooses its new velocity from the set of permitted velocities that do not result in collision with other agents (line 17). This is done as described in equation (\ref{newvelocity}).

\begin{algorithm}[t]
    \caption{Computation of Collision-Free Velocity}
    \begin{algorithmic}[1]
        \REQUIRE $X$
        \ENSURE $v_{i,j}^{new}$
            \STATE $X \leftarrow$ update($X$);
            \STATE $x_{i,j} \leftarrow X_j(i)$;
            \STATE $Edges_j,Vertices_j \leftarrow$ voronoi($X_j,Q$);
            \STATE $CM_{V_{i,j}} \leftarrow$ centerOfMass($Edges_j(i),Vertices_j(i)$);
            \STATE $v_{i,j}^{pref} \leftarrow$ desiredVelocity($CM_{V_{i,j}},x_{i,j},c_{i,j}$);
            \FORALL{$X_l \subset X, \, l=\{1,...,N\} $} 
                \FORALL{$x_{k,l} \in X_l, \forall \, (k,l) \neq (i,j),k=\{1,...,m\}$}
                    \STATE $Edges_l,Vertices_l \leftarrow$ voronoi($X_l,Q$);
                    \STATE $CM_{V_{k,l}} \leftarrow$ centerOfMass($Edges_l(k),Vertices_l(k)$);
                    \STATE $\hat{v}_{k,l}^{pref} \leftarrow$ desiredVelocity($CM_{V_{k,l}},x_{k,l},\hat{c}_{k,l}$);
                    \STATE $\mathcal{VO}_{i,j|k,l}^{\tau} \leftarrow$ velocityObstacle($x_{i,j},x_{k,l},r_{i,j},r_{k,l},\tau$);
                    \STATE $ORCA_{i,j|k,l}^{\tau} \leftarrow$ halfplane($\mathcal{VO}_{i,j|k,l}^{\tau},v_{i,j}^{pref},\hat{v}_{k,l}^{pref}$);
                \ENDFOR
                \STATE $ORCA_{i,j|l}^{\tau} \leftarrow \bigcap_{k=1}^{m} ORCA_{i,j|k,l}^{\tau}$;
            \ENDFOR
            \STATE $ORCA_{i,j}^{\tau} \leftarrow D(0,v_{i,j}^{max}) \cap  \bigcap_{l=1}^{N-1} ORCA_{i,j|l}^{\tau}$;
            \STATE $v_{i,j}^{new} \leftarrow$ newVelocity($ORCA_{i,j}^{\tau}$);
        \RETURN{} $v_{i,j}^{new}$
     \end{algorithmic}
     \label{algorithm:v_new}
\end{algorithm}

\section{Simulation and Results} \label{simulation}

To validate the functionality of the algorithm proposed in section \ref{proposed method}, this section presents results obtained through Monte Carlo simulations of multiple non-cooperating swarms independently covering a common area followed by a specific example. All simulations are performed in Matlab and consider the same area as example \ref{ex:example1} with the vertices of the boundaries being located at $[(1,0),(6,0),(8,5),(5,8),(0,4)]$ and two swarms each consisting of four agents. For collision avoidance, the time horizon $\tau$ is chosen to be one iteration which implies that the computed $\mathcal{VO}$s include the velocities that result in collision at the following iteration.

\begin{figure}[t]
    \centering
    \includegraphics[scale=0.5]{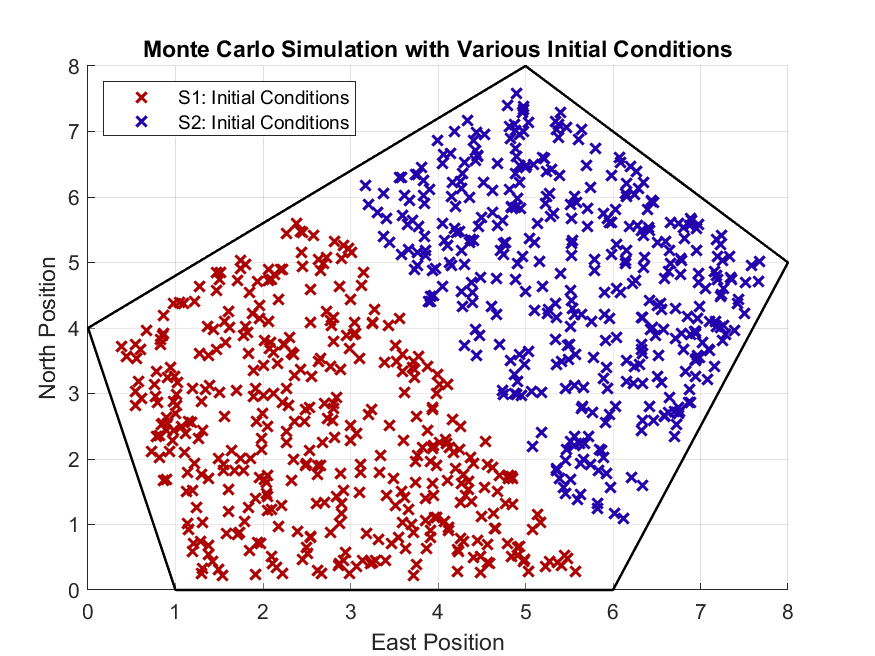}
    \caption{Initial positions of agents tested in Monte Carlo simulations}
    \label{fig:montecarlo}
\end{figure}

Monte Carlo simulations of 100 iterations are executed to demonstrate that the agents using the proposed algorithm do not collide. For each iteration, the initial locations of the four agents of each swarm are chosen randomly within two different subsets of the area under consideration and displayed in figure \ref{fig:montecarlo}. Distinct swarms always starting in different subsets of the area results in comparatively long trajectories. Therefore, increased encounters with agents from the other swarm are more likely, since it assures that in the beginning part of the area is not well covered. Notice that each agent always starts with the minimum distance from each other agent as well as the boundaries of the area. This distance is the sum of the agents own radius and that of another vehicle. For 100 computations of motion of the two swarms to centroidal Voronoi configurations no collision occurred. A specific example demonstrating one iteration of the Monte Carlo simulation follows in example \ref{ex:example2}.

\begin{figure*} [t]
    \centering
    \subfloat[]{\includegraphics[width=0.33\textwidth]{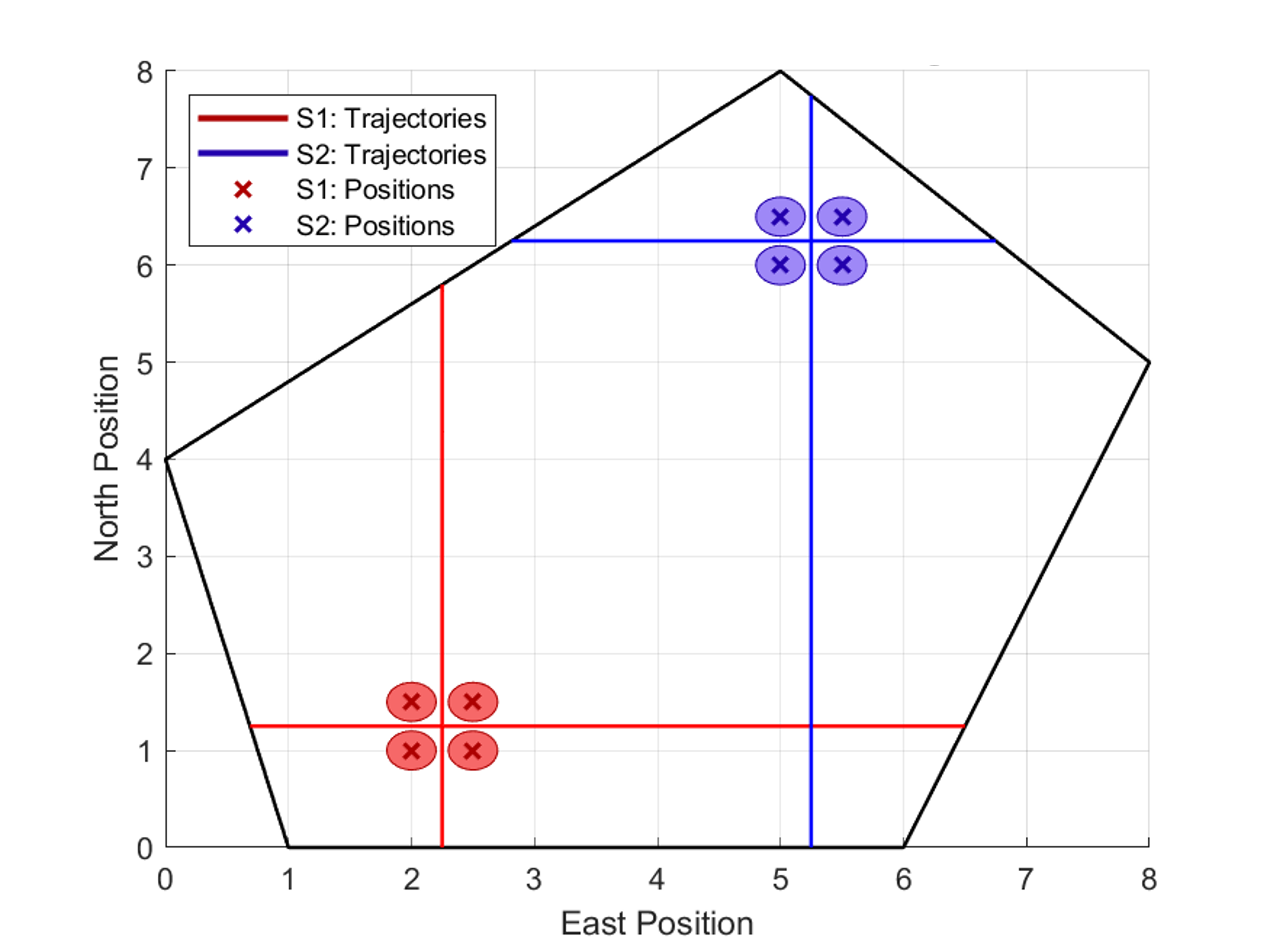}}
    \subfloat[]{\includegraphics[width=0.33\textwidth]{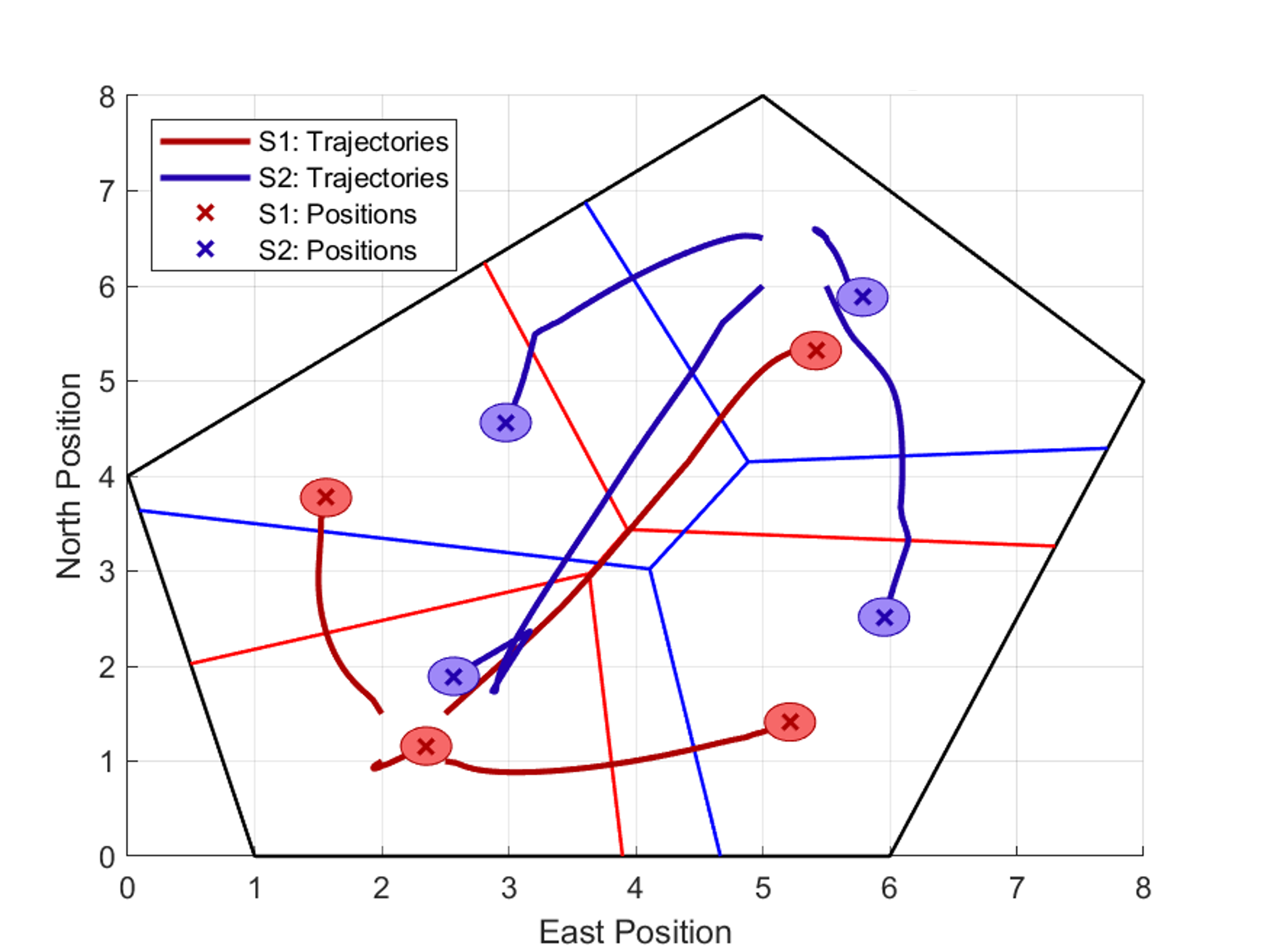}}
    \subfloat[]{\includegraphics[width=0.33\textwidth]{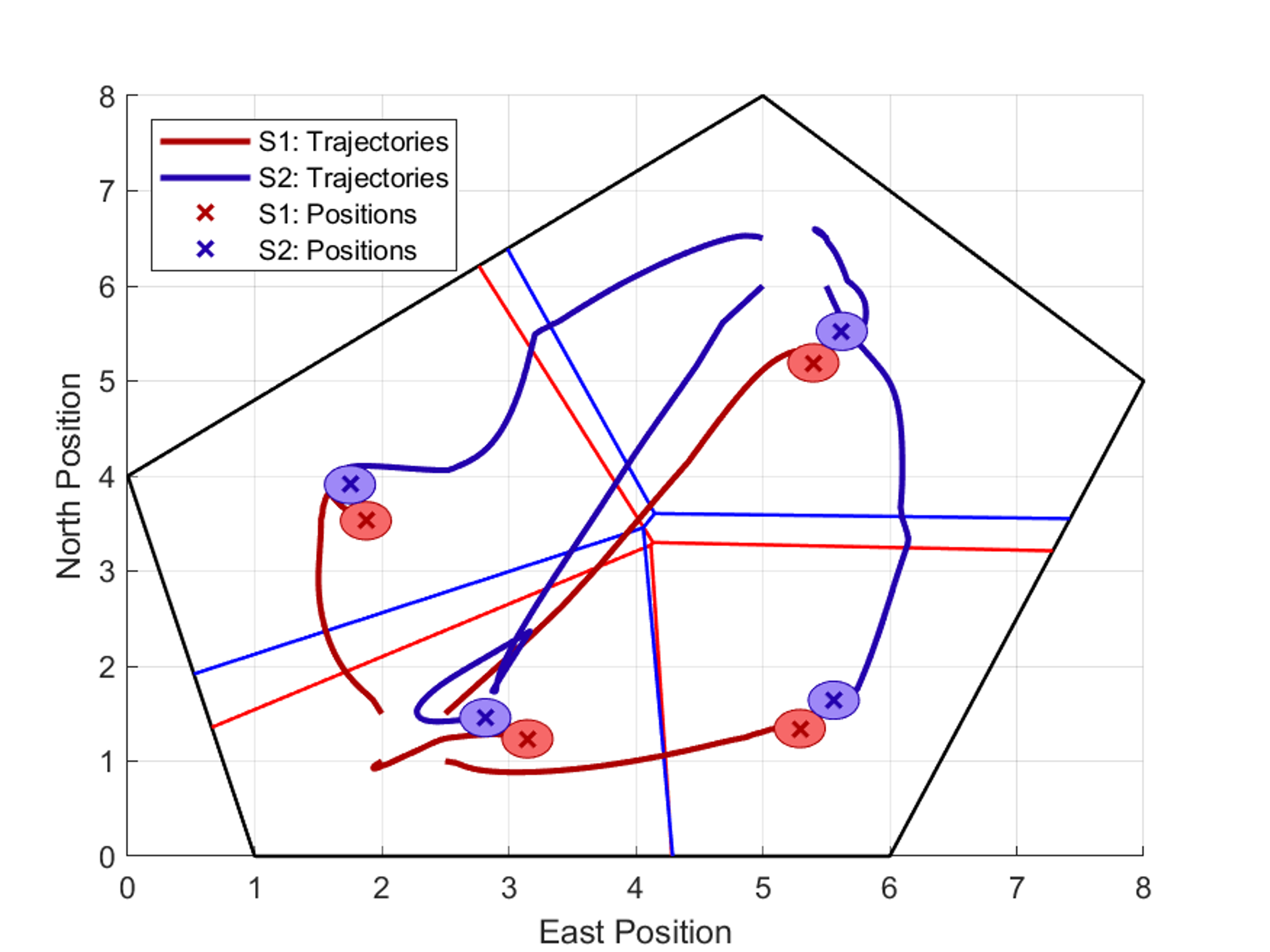}}

    \caption{Simulation of coverage control with two non-cooperating swarms with collision avoidance showing the trajectories and final positions of the agents of swarm 1 (S1) and swarm 2 (S2) (a) at the first iteration (b) after 50 iterations (c) after convergence to their final positions.}
    \label{fig:simulation}
\end{figure*}

\begin{example} \label{ex:example2}
This example illustrates the coverage task executed by distinct swarms with different initial positions within a common area using the proposed algorithm. Collision avoidance is therefore implemented. Supposing the same conditions as in example \ref{ex:example1}, the density distribution is uniform. Again, two swarms each containing four agents are placed at the following positions:
 \begin{itemize}
     \item swarm 1 (S1): $[(2,1),(2,1.5),(2.5,1),(2.5,1.5)]$,
     \item swarm 2 (S2): $[(5,6),(5,6.5),(5.5,6),(5.5,6.5)]$.
 \end{itemize}
The radii of all agents are $0.2$ units. 

The obtained results are depicted in figure \ref{fig:simulation}. The agents from both groups are shown at their initial locations in figure \ref{fig:simulation}a. After the first 50 iterations (figure \ref{fig:simulation}b), the trajectories traveled by the agents are still similar to those observed after the same amount of iterations in example \ref{ex:example1}. Notice that despite trajectories seeming to cross each other, no collisions occurred, since agents reached these intersections at different points in time. As the agents get closer to their desired final positions (figure \ref{fig:simulation}c), differences to the simulation without obstacle avoidance in example \ref{ex:example1} are clearly visible. Instead of reaching the centers of mass of their Voronoi cells, agents from different swarms converging to the same final position compete for the respective spots to achieve maximum coverage. Yet, collisions do not occur which supports the suitability of the proposed algorithm for safe motion.
    
\end{example}

\section{Conclusions} \label{conclusions}

This paper presented a methodology for multiple non-cooperating swarms of UAVs to independently cover a common area. An algorithm suitable for collision avoidance in coverage control with distinct groups of agents was proposed and applied in various examples. Despite agents encountering each other during the pursuance of the coverage task and the occurrence of pairs of agents from different swarms competing to move to their desired location, extensive Monte Carlo simulations show that the proposed algorithm ensures collision-free motion. Future work includes the incorporation of uncertainty in the estimates of the agents' desired velocities as well as the extension to three-dimensional scenarios. The incorporation of non-uniform density distributions over the area to be covered is another possibility.

\addtolength{\textheight}{-12cm}   


\bibliographystyle{IEEEtran}
\bibliography{IEEEabrv,References}

\end{document}